\documentclass[letterpaper]{article} 
\usepackage{aaai25}  
\usepackage{times}  
\usepackage{helvet}  
\usepackage{courier}  
\usepackage[hyphens]{url}  
\usepackage{graphicx} 
\usepackage{natbib}  
\usepackage{caption} 
\usepackage{latexsym}
\usepackage{amssymb}
\usepackage{amsmath}
\usepackage{amsthm}
\usepackage{booktabs}
\usepackage{enumitem}
\usepackage{graphicx}
\usepackage{calc}
\usepackage{tikz}
\usetikzlibrary{automata, calc, positioning,arrows.meta}
\usepackage[linesnumbered,ruled,vlined]{algorithm2e}

\newtheorem{theorem}{Theorem}

\newtheorem{definition}{Definition}

\newcommand{\BibTeX}{B\kern-.05em{\sc i\kern-.025em b}\kern-.08em\TeX}

\usepackage{listings}
\usepackage{xcolor}

\definecolor{codegray}{rgb}{0.5,0.5,0.5}
\definecolor{backcolour}{rgb}{0.99,0.99,0.99}
\definecolor{commentgreen}{rgb}{0,0.6,0}
\definecolor{keywordblue}{rgb}{0,0,0.8}
\definecolor{stringred}{rgb}{0.6,0,0}

\title{Declarative Integration and Management of Large Language Models through Finite Automata: Application to Automation, Communication,  and Ethics}
\author{
    Thierry Petit$^{1,2}$, Arnault Pachot$^{1}$, Claire Conan-Vrinat$^{1}$, Alexandre Dubarry$^{1}$ 
}
\affiliations{
     \textsuperscript{$^1$EMOTIA, 59, rue Boissière 75116 Paris, France.} \\ 
     \textsuperscript{\texttt{contact@emotia.com}} \\
     \textsuperscript{$^2$Worcester Polytechnic Institute, 100 Institute Rd, Worcester, MA 01609, USA.} \\ 
     \textsuperscript{\texttt{tpetit@wpi.edu, tpetit19@gmail.com}} 

}

\begin{document}

\maketitle








\author{}





\begin{abstract}
This article introduces an innovative architecture designed to declaratively combine Large Language Models (LLMs) with shared histories, and triggers to identify the most appropriate LLM for a given task. Our approach is general and declarative, relying on the construction of finite automata coupled with an event management system. The developed tool is crafted to facilitate the efficient and complex integration of LLMs with minimal programming effort, especially, but not only, for integrating methods of positive psychology to AI. The flexibility of our technique is demonstrated through applied examples in automation, communication, and ethics. 
\end{abstract}



\section{Introduction}
Large Language Models (LLMs) have heralded a revolution in the field of human-computer interaction, primarily due to their ability to simulate human behaviors~\cite{aher2023using}. 
In the industrial context, however, integrating LLMs to an application requires significant resources in development and validation.
The inherently stochastic process of response generation poses ethical and operational considerations. 
To circumvent this issue, it is now increasingly recognized that chaining multiple LLMs can yield more reliable outcomes than relying on a single LLM~\cite{zheng2023progressivehint,sreedhar2024simulating,zeng2021enhancing,wei2022chain}. 
Several libraries have emerged to facilitate the chaining of Large Language Models (LLMs), e.g., \texttt{langchain}~\cite{Chase2022LangChain}. 
Chaining LLMs requires developing a specific model for each application, addressing the issue of detection that determines the dynamic sequencing of LLMs, validation, and maintaining a response time compatible with the context of use.

In this article, we present a novel approach aimed at integrating LLMs and other AI modules through a \textit{declarative} method, eliminating the need, as much as possible, for challenging programming tasks.
Unlike LLMStack~\cite{llmstack}, which primarily facilitates the chaining of LLMs for application generation, our approach employs a \emph{model} to manage multi-modal interactions transparently. This enables the integration of triggers and subtle conversation history sharing mechanisms, without presupposing specific applications.
With our framework, designing a multi-agent system based on LLMs involves the following steps:
    (1) Define an \textit{automaton} whose states are LLMs/AI modules dedicated to different tasks, specifying the sharing of history.
    (2) Define the edges as \textit{triggers}, evaluating the conditions to transition from one state to another.
    (3) Set trigger \textit{priorities} when a state can have multiple successors.
    (4) Write the LLMs' prompts.
The platform manages the sequential triggering of the states (e.g., LLMs) in the automaton. It ensures the updating of history shared between subsets of states and arcs. 
Using our approach, one may quickly test several models for an application, without the need for procedural development or algorithms to implement each trial. 

We introduce our framework by considering dialogues between humans and machines. However, it also supports the integration of problem-solving or any other AI modules provided that they implement an interface that allows the system to transparently manage interactions between states of the automaton and the history. We can use diverse triggers such as visual detection, sound, onboard sensors, or any other form of data. 
Thereby, although there exists links between our framework and AI planning tools for dialogue management (such as DRUID AI, or Rezolve.ai), our theoretical framework is unique, due to  minimal assumptions on automata states and transitions.
To demonstrate the flexibility of our technique, we present three examples based on LLMs: an example of automating train ticket booking, a non-violent communication scheme, and an example related to prevent ethical issues with LLMs, which are especially challenging in multi-modal systems~\cite{sutton2018reinforcement}. 

\section{Multi-Modal Models Based on Automata}
We aim to ensure the three following characteristics:
1. The chaining of LLMs and other AI modules\footnote{Without loss of generality, we will consider LLMs in this section, but any AI modules exchanging data can be considered.} is decribed by a model, independently of its implementation, including the triggers for determining the order of use of the different LLMs.
2. The sharing of the conversation history is described declaratively and managed transparently for the user.
3. The response time can be estimated from the structure.

We define an automaton whose states are LLMs or user's messages, and transitions are triggers 
associated with a priority to determine which LLM is to be invoked next.

\begin{definition}[state]\label{def:state}
Given an alphabet \(\Sigma\), a \textbf{state} in an automaton is associated with a function 
\(q : \Sigma^* \rightarrow \Sigma^+\), where \(\Sigma^*\) is the set of all possible strings (including the empty string) that can be formed from the alphabet \(\Sigma\) and \(\Sigma^+\) denotes the set of all non-empty strings.
\end{definition}

\begin{definition}[final state]\label{def:accept_state}
A \textbf{final state} is a specific type of state for which the output string \(q(s)\) for any input string \(s \in \Sigma^*\) satisfies a condition where the message exchange can conclusively end. 
\end{definition}

We consider both a user and an LLM as functions that map messages to messages. We associate these functions with states in a finite-state machine. A state associated with a user message is inherently a final state, signifying the user's freedom to end the conversation. States related to LLMs are rarely final, though we do not restrict this possibility. 

\begin{definition}[MFA]\label{def:mfa}
    Given an alphabet \(\Sigma\), a \textbf{Multi-LLM Finite Automaton (MFA)} is a 4-tuple, $(Q, \delta, q_0, F)$:
    A finite set of states $Q$, which can be LLMs, AI modules or user messages.
    A transition function $\delta: Q \times \Sigma \times \mathbb{N}^{|Q|} \rightarrow Q$, where $\mathbb{N}^{|Q|}$ represents a list of positive integers of size $|Q|$. 
    An initial state $q_0 \in Q$.
    A set of final states $F \subseteq Q$.
\end{definition}
The construction of the transition function is based on triggers. 
A trigger is a module that determines whether a state (LLM or awaiting user message) should be activated or not, from the output sent by another state. Therefore, a trigger is an \textit{arc} in the MFA that enables composing the transition function of the automaton. The trigger will respond with an integer representing a Boolean (a bit): 0 indicates that the candidate should not be activated, and 1 signifies acceptance. Given that there may be multiple candidate states to continue the interaction, the user will determine an integer priority in $\{1, \ldots, |Q|\}$ among all the candidates outgoing each state. When a state is not a candidate, its priority is $0$. 
If two triggers answer positively at the output of a state, the one with the highest priority prevails: the next state will be the one corresponding to this arc (transition).

\begin{definition}[Trigger]\label{def:trigger}
Given a MFA  $(Q, \delta, q_0, F)$, a state $q \in Q$, a string $s$ and a priority $p \in \{0, \ldots, |Q|\}$, let's define 
$f_\tau(q,s) :  Q \times \Sigma^* \rightarrow \{0,1\}$, a function that assigns a binary value based on the given state and message.\footnote{We add $q$ for generality: $f_\tau(q,s)$ is often independent of $q$.}
A \textbf{trigger} is a  function $\tau : Q \times \Sigma^* \times p \rightarrow \{0,1\}$ defined as follows: 
 $\tau(q,s,p) = \mathit{min(p\times f_\tau(q,s),p)}$. 
The set of triggers is denoted by $T$. 
\end{definition}
While it may seem logical for the priorities associated with transitions leaving a state to establish a total order, we do not impose this restriction. If the model designer defines multiple priorities as equal, and at a certain stage of the interaction multiple equivalent candidates are accepted, we suggest randomly selecting the next state. In case of a dialogue, this feature can help to simulate naturally unpredictable parts.

The last step consists of handling how dialogue history is shared by the LLMs. 
For this purpose, we define a bipartite graph linking states and triggers\footnote{Triggers may be stated as LLMs.} to computational objects representing histories. 

\begin{definition}[History Graph]\label{def:history_graph}
Let $M = (Q, \delta, q_0, F)$ be an MFA with triggers $T$. A \textbf{history} is a non-empty set of sequences in an alphabet $\Sigma^*$.
The set of histories is denoted by $H$.
The bipartite history graph $\mathcal{H} = ((Q \cup T,H),E=E_{rw}\cup E_r \cup E_w)$ defines the history attachment. $\forall x \in Q \cup T$, there is at most one edge $e \in E$ linking $x$ to a vertex $h \in H$. Such an edge indicates that $x$ is attached to $h$. 
    For $e=(x,h) \in E_{rw}$, $x$ reads and updates $h$.
    For $e=(x,h) \in E_{r}$, $x$ only reads $h$.
    For $e=(x,h) \in E_{w}$, $x$ only updates $h$.

\end{definition}
The workload can be computed from the automaton by estimating the (maximum) number of states between two user messages and considering the average processing time of each state. 

\begin{algorithm}[h!]\scriptsize
\caption{MFA-based dialogue}
\label{alg:pseudo}
\SetAlgoNoLine 
\SetAlgoNoEnd 

$c \gets$ initial state of the MFA\; 
$s \gets$ "~"\; 
\While{continue the chat}{ 
    \eIf{$c$ is a user node}{
        $s \gets$ get input from the user\; 
    }{
        $r \gets$ output from $c$ state\; 
        Add $s$ and $r$ to $c$ (shared) history\;
    }
    ${T} \gets$ list of triggers adjacent to $c$ returning 1\; 
    $c \gets$ endpoint of an edge $\tau \in {T}$ with maximum priority\;
}
\end{algorithm}

Algorithm~\ref{alg:pseudo} outlines the basic structure of a dialogue automatically derived from any MFA.

By using this framework, if the states and transitions of the MFA are based on LLMs, 
no specific computer skills are required. The steps are as follows: write the prompts for the LLMs, including some triggers, define the MFA that models the chronological sequencing of the states (with transition priorities and terminal states).

\paragraph{Example.}
To illustrate our framework, we use a communication method derived from the DESC (Describe, Express, Suggest, and Conclude) method~\cite{bower2009asserting} and related to handling client complaints. This method is tailored, 
for example, for welcoming quests in a restaurant.
\begin{figure}[h!]
\centering
\begin{tikzpicture}[shorten >=1pt, node distance=2.2cm, on grid, auto, state/.style={draw, circle, minimum size=7pt}]
    \node[state] (q1) {$l_1$};
    \node[state, accepting, right=of q1] (q0) {$q_0$};
    \node[state] (q2) [below=of q0, yshift=0.9cm] {$l_2$};
    \node[state] (q3) [right=of q2, accepting] {$q_3$};
    \node[state] (q4) [right=of q3] {$l_4$};
    \path[->]
    (q0) edge [bend left] node {$t_1, 1$} (q1)
    (q1) edge [bend left] node {$t_1, 1$} (q0)
    (q0) edge node {$t_0, 2$} (q2)
    (q2) edge node {$t_1, 1$} (q3)
    (q3) edge node {$t_1, 1$} (q4)
    (q4) edge [bend right] node {$t_1, 1$} (q0);
\end{tikzpicture}
\caption{\small MFA of the ARPS technique. $q_0$: User message (start and final).
$l_1$: Standard LLM.
$l_2$: LLM for acknowledging the client's complaint, reformulating, and probing.
$q_3$: User message (final).
$l_4$: LLM for suggesting solutions. Arcs are labelled with trigger name and priority.} 
\label{fig:example}
\end{figure}
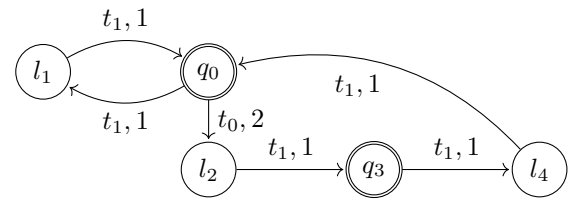

We introduce the ARPS technique, consisting of  four steps: (1) Acknowledge the client complaint. (2) Rephrase it. 
(3) Probe to show interest and eventually obtain more details about the complaint. (4) Suggest a solution. 

The MFA stated in Figure~\ref{fig:example} is based on two triggers:
    $t_0$: detects if the client is complaining. $f_{t_0}$ can be, for instance, implemented with a dedicated LLM prompted to detect anger or frustration. We associate the priority $p_{t_0}=2$ to this trigger. 
    $t_1$: $\forall m \in \Sigma^*$, $f_{t_1}(m) = 1$, with priority $p_{t_1}=1$.

The states $Q$ are the following: 
$q_0$: User message.
$l_1$: Standard LLM. 
$l_2$: LLM for acknowledging, reformulating, and probing.
$q_3$: User message in the context of ARPS.
$l_4$: LLM for suggesting solutions to clients' complaints.

There is only one history $h$, which, chronologically, keeps records of the user's messages and answers of LLMs. The graph $\mathcal{H} = ((Q,H),E)$ is such that
$E = E_{rw} = \{(l_1, h), (l_2, h), (l_4,h)\}$.

The MFA encodes a scenario where the dialogue continues in a standard manner between a user ($q_0$) and an LLM ($l_1$), unless the trigger $t_0$ detects a message of anger or complaint. If so, the LLMs representing the ARPS pattern take over. Once the conflict is resolved, the dialogue returns to standard. This example is simplified, as it considers that only one interaction allows us to solve the conflict. We study a more sophisticated model in the 'Case Studies' Section.


\newlength{\speakerwidth}
\settowidth{\speakerwidth}{Dialer} 

\begin{table}[h!]
\scriptsize 
\centering
\begin{tabular}{@{}p{\speakerwidth + 2\tabcolsep}p{\columnwidth - \speakerwidth - 4\tabcolsep}@{}}
\toprule
\textbf{Speaker} & \textbf{Message} \\
\midrule
User & hello \\
\addlinespace
Dialer & Hello! How are you today? \\
\addlinespace
User & It’s outrageous to take half an hour to serve a sandwich! \\
\addlinespace
Dialer & It certainly can be frustrating to wait for an extended period, especially for something as seemingly straightforward as a sandwich. Long wait times at restaurants or food establishments can result from various factors, such as a high volume of customers ... \\
\bottomrule
\end{tabular}
\caption{\small Dialogue between User and Dialer without ARPS.}\label{tab:1}
\end{table}

\begin{table}[h!]
\scriptsize 
\centering
\begin{tabular}{@{}p{0.5cm}p{\speakerwidth + 2\tabcolsep - 0.5cm}p{\columnwidth - \speakerwidth - 4\tabcolsep - 0.5cm}@{}}
\toprule
\textbf{MFA state} & \textbf{Speaker} & \textbf{Message} \\
\midrule
$q_0$ & User & hello \\
\addlinespace
$l_1$ & Dialer & Hello! How are you today? \\
\addlinespace
$q_0$ & User & It’s outrageous to take half an hour to serve a sandwich! \\
\addlinespace
$l_2$ & Dialer & I understand that you are frustrated with the long wait time for your sandwich. Can you tell me more about this issue? \\
\addlinespace
$q_3$ & User & I have to go back to work quickly! \\
\addlinespace
$l_4$ & Dialer & We will suggest implementing a pre-made sandwich option to reduce wait time for customers in a hurry. \\
\bottomrule
\end{tabular}
\caption{\small Dialogue between User and Dialer with ARPS.}\label{tab:2}
\end{table}
Tables~\ref{tab:1} and \ref{tab:2} compare a dialogue produced using this MFA to a dialogue with a single LLM, which would correspond to an MFA reduced to the two states $q_0$ and $l_1$. We used OpenAI's ChatGPT 3.5~\cite{openai2021chatgpt} for this example, with an unprompted standard LLM $l_1$. 
Since the example is based solely on LLMs (dialers and triggers), our task involved only providing the automaton in Figure 1 as input data and writing prompts (anger detection and ARPS). 
\paragraph{Engineering considerations.}
The system implementation requires ensuring two main points:
\begin{enumerate}
  \item \textit{Generality}: the system must ensure that there are no restrictions on the LLMs/modules and triggers used.
  \item \textit{Transparently shared history}: to be fully declarative, the system must allow invisible data management once histories are attached to the graph components.
\end{enumerate}

To address the first point, the states and arcs of the MFA are wrappers of objects constrained to implement specific interfaces. We employ wrappers since both states and edges may be represented as LLMs and have shared histories, as outlined in Definition~\ref{def:history_graph}.
The interface for LLMs includes the following functions: \texttt{predict(user\_message)}, which returns the LLM's response based on a user message, and \texttt{add(inputm, outputm)}, which adds a (message, response) pair to the history if the LLM is attached to one (otherwise, it has no effect). Triggers, which can be LLMs, should not write to the history, though they can read it. They have a priority, so the basic interface includes \texttt{predict(user\_message)}, returning the trigger's \{0,1\} response from a message, \texttt{get\_priority()}, which returns the integer priority, and \texttt{set\_priority(p)}, which assigns the integer priority to \texttt{p}.

The second point requires event-based programming.
The history is shared using the Observer/Observable design pattern~\cite{Gamma1994}. The \texttt{Archive} class is Observable and 
includes an \texttt{add(inputm, outputm)} method (similar to the one of the LLM interface), and methods for returning and  removing (message, response) pairs. This class notifies all its observers at each new event, such as adding or removing data. The \texttt{History} class connects a specific LLM/Module and an \texttt{Archive} object, using an \texttt{update} method designed to respond to events within its archive, including its own addition of a (message, response) pair.

The \texttt{LLM} class encapsulates a \texttt{History} object linked to a specific \texttt{Archive} reference. Any MFA state that is an LLM should inherit from the \texttt{LLM} class. If not, it must at least adhere to the LLM interface. MFA edges must conform to the trigger interface and can be designed as subclasses of the \texttt{LLM} class, depending on whether a LLM is utilized to implement the trigger.
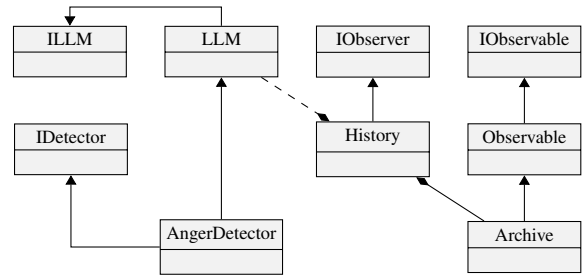
\begin{figure}[h]
\centering
\scriptsize
\tikzset{
    classnode/.style={
        draw,
        rectangle split,
        rectangle split parts=2,
        text centered,
        minimum width=1.5cm,
        fill=gray!10
    },
    inherit/.style={->, >=Triangle},
    composition/.style={->, >=Diamond},
    aggregation/.style={->, >=Diamond, dashed},
    line/.style={-}
}

\begin{tikzpicture}[node distance=1.3cm and 0.5cm]

\node[classnode] (llm)
{
    LLM
};

\node[classnode, left=0.5cm of llm] (illm)
{
    ILLM
};

\node[classnode, right=0.5cm of llm] (iobserver)
{
    IObserver
};

\node[classnode, below of=iobserver] (history)
{
    History
};

\node[classnode, right=0.5cm of iobserver] (iobservable)
{
    IObservable
};

\node[classnode, below of=iobservable] (observable)
{
    Observable
};

\node[classnode, below of=observable] (archive)
{
    Archive
};

\node[classnode, below of=llm, xshift=-2cm] (idetector)
{
    IDetector
};

\node[classnode, below of=idetector, xshift=2cm] (angerdetector)
{
    AngerDetector
};

\draw[inherit] (llm.north) -- ++(0,0.25) -| (illm.north);
\draw[inherit] (observable.north) -- ++(0,0.5) -| (iobservable.south);
\draw[inherit] (history.north) -- ++(0,0.5) -| (iobserver.south);
\draw[inherit] (angerdetector.west) -| (idetector.south);
\draw[inherit] (angerdetector.north) -| (llm.south);
\draw[inherit] (archive.north) -| (observable.south);

\draw[aggregation] (llm) -- (history);
\draw[composition] (archive) -- (history);

\end{tikzpicture}\caption{\small Event-based architecture for shared history with \texttt{AngerDetector}, a specific LLM-based trigger.} 
\end{figure}
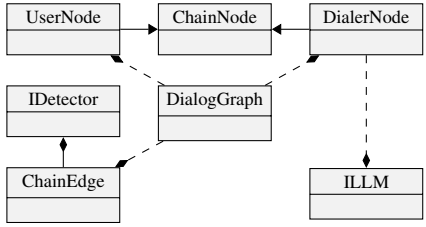
\begin{figure}[h]
\centering
\scriptsize
\tikzset{
    classnode/.style={
        draw,
        rectangle split,
        rectangle split parts=2,
        text centered,
        minimum width=1.5cm,
        fill=gray!10
    },
    inherit/.style={->, >=Triangle},
    composition/.style={->, >=Diamond},
    aggregation/.style={->, >=Diamond, dashed},
    line/.style={-}
}
\begin{tikzpicture}[node distance=0.4cm and 0.5cm]

\node[classnode] (chainnode) {ChainNode};
\node[classnode, right=of chainnode] (dialernode) {DialerNode};
\node[classnode, left=of chainnode] (usernode) {UserNode};
\node[classnode, below=of usernode] (idetector) {IDetector};
\node[classnode, below=of idetector] (chainedge) {ChainEdge};
\node[classnode, below=of chainnode] (dialoggraph) {DialogGraph};
\node[classnode, below=1.5cm of dialernode] (illm) {ILLM}; 

\draw[inherit] (dialernode) -- (chainnode);
\draw[inherit] (usernode) -- (chainnode);

\draw[aggregation] (dialoggraph) -- (dialernode);
\draw[aggregation] (dialoggraph) -- (usernode);
\draw[aggregation] (dialoggraph) -- (chainedge);
\draw[aggregation] (dialernode) -- (illm);

\draw[composition] (chainedge) -- (idetector);

\end{tikzpicture}
\caption{\small Class hierarchy for MFA representation and use.}
\end{figure}
At last, we define wrapping classes for easily maintain the graph representing the MFA and to implement generic exploration procedures (such as Algorithm~\ref{alg:pseudo}). The \texttt{ChainNode} class defines a state of the MFA and allows specifying how it interacts with its history ($r, w,$ or $rw$). This class has two subclasses, distinguishing states where the machine responds, \texttt{DialerNode}, and those where the user enters a message, \texttt{UserNode}. The \texttt{DialerNode} class possesses a method that indicates whether the LLM should display its response to the user or not (based on its position in the MFA and its role). The \texttt{ChainEdge} class represents the transitions. It is in this class that the list of triggers corresponding to an originating state is stored. The \texttt{DialogGraph} class contains methods to build and explore the automaton. 


The platform was implemented in Python. For Example~\ref{fig:example}, the model is as follows:
\begin{lstlisting}
model = "gpt-3.5-turbo"
q0 = UserNode()    
dialer1 = Dialer(...)              # context prompt
l1 = DialerNode(dialer1)
dialer2 = Dialer(...)              # prompted for A,R,P
l2 = DialerNode(dialer2)
q3 = UserNode()
dialer4 = Dialer(...)              # prompted for S
l4 = DialerNode(dialer4)           
q0l1 = ChainEdge()                 # t1       
detector_anger = AngerDetector(model)
detector_anger.set_priority(2)
q0l2 = ChainEdge([detector_anger]) # t0
l1q0 = ChainEdge()                 # t1
... # similar for l2q3, q3l4 and l4q0 (t1)
mfa = Automaton()
mfa.add_state(q0)          
... # similar for all the states
mfa.add_edge(q0l1)
... # similar for all the edges
mfa.run()                          # e.g., Algorithm 1
\end{lstlisting}

\paragraph{Triggers.}
For an application to function, the triggers must be reliable and provide their responses quickly. 
For example, for NVC, the effectiveness of LLMs hinges on their ability to subtly interpret the emotional state of the user. 
We propose a simple and systematic test protocol. Let us first consider a trigger dedicated to conversational analysis. The protocol is based on three phases: 
\begin{enumerate}
\item \textit{Sentence Generation}: Sentences are generated either manually or using a Language Model (LLM) with a specific goal corresponding to a trigger, expressed as a simple zéro-shot prompt. For instance, \textit{"generate sentences expressing anger or frustration in the context where a restaurant customer complains about their experience."} When an LLM is used for generating sentences, they must undergo human validation. 
    \item \textit{Dataset Augmentation}: Randomly generated sentences (potentially unrelated to the trigger) are incorporated into the dataset. This addition diversifies the dataset. 
    \item \textit{Human Expert Validation}: \textit{Human experts} manually validate the dataset by assigning a binary value (0 or 1) to each sentence, indicating its alignment with the trigger under consideration. 
    \item \textit{Testing and Evaluation}: Test the trigger on the dataset by comparing its assignments with the manual assignments and evaluating execution time. 
\end{enumerate}
In a more general context, e.g., a MFA involving triggers which are not LLMs, and/or when entry data are not sentences in natural language, the general philosophy of the protocol is the same, except the validation phase, which is not required to be performed by humans. The validation can be conducted using algorithms more sophisticated and time-consuming than the trigger. 
For dialogue systems, we empirically observed that being above 75\% of positive inferences was enough for improving the results obtained with a single LLM, and that the best temperature for triggers is 0.1. 
\section{Case Studies}
\paragraph{Automated Train Ticket Booking.}
Integrating multiple LLMs with specialized modules for automated tasks enhances the accuracy and standardization of data collection from user dialogues. 
Let's consider a "natural language" train ticket reservations example that demonstrates the ease of modeling with our approach. It is kept minimalistic to avoid unnecessary complexity.
The idea is to query for inputs, looping until the response can be entered into the system that requires formatted data. 
Some states are writer modules, i.e., functions writing an input in a database or CSV file, and returning no output. 
The triggers are the following: 
    $t_0$: City name trigger. $f_{t_0}$ is, for instance, tasked with verifying whether the user's message includes an actual city name (possibly in a specific list). The priority level is $p_{t_0}=2$.
    $t_1$: Time trigger: The function $f_{t_1}$, for example, may be designed to ascertain if the user's message contains a time specification (return 1) or not (return 0), with priority $p_{t_1}=2$.
    $t_2$: $\forall m \in \Sigma^*$, $t_2$: $f_{t_2} = 1$, with priority $p_{t_2}=1$.

The states $Q$ are the following: 
$q_0$: User message.
$l_1$: LLM for departure city inquiry. While a simple print statement could serve to ask for the departure city, employing an LLM allows for varied question phrasing. This flexibility is beneficial for repeated inquiries following unclear user responses. Therfore, this LLM must share the global history. 
$w_2$: 
Writer module: add the city name to a database/CSV file.  
$l_3$: 
LLM for destination city request.
$q_4$: 
User message.
$w_5$: 
Writer module: add the city name to a database/CSV file. 
$l_6$: 
LLM for departure time inquiry.
$q_7$: 
User message.
$l_8$: 
LLM used to extract and convert the time in a standard format.  
$w_9$: 
Writer module: add the time to a database/CSV file. 

The history $h$ be shared by all states corresponding to LLMs, excluding user messages and writers, to keep track of the exchanges at each stage. In this example, it is not mandatory to share this history with triggers (they only use the last output). 
The graph $\mathcal{H} = ((Q,H),E)$ is such that
$E = E_{rw} = \{(l_1, h), (l_3, h), (l_6,h), (l_8,h)\}$.

\begin{figure}
\centering
\begin{tikzpicture}[shorten >=1pt, node distance=2.2cm, on grid, auto, state/.style={draw, circle, minimum size=7pt}]

    \node[state] (q1) {$l_1$};
    \node[state, accepting] (q0) [right=of q1] {$q_0$};
    \node[state] (q2) [below=of q0, yshift=0.7cm] {$w_2$};
    \node[state] (q3) [right=of q0] {$l_3$};
    \node[state, accepting] (q4) [right=of q3] {$q_4$};
    \node[state] (q6) [below=of q3, yshift=0.7cm] {$l_6$};
    \node[state] (q5) [right=of q6] {$w_5$};
    \node[state,accepting] (q7) [below=of q2,, yshift=0.7cm] {$q_7$};
    \node[state, accepting] (q9) [below=of q1, yshift=0.7cm]{$w_9$};
    \node[state] (q8) [below=of q9, yshift=0.7cm]{$l_8$};

    \path[->]
    (q0) edge [bend left] node {$t_2, 1$} (q1)
    (q1) edge [bend left] node {$t_2, 1$} (q0)
    (q0) edge [bend right] node {$t_0, 2$} (q2)
    (q2) edge [bend right] node {$t_2, 1$} (q3)
    (q3) edge [bend right] node {$t_2, 1$} (q4)
    (q4) edge [bend right] node {$t_2, 1$} (q3)
    (q4) edge node {$t_0, 2$} (q5)
    (q5) edge node {$t_2, 1$} (q6)
    (q6) edge [bend left] node {$t_2, 1$} (q7)
    (q7) edge  node {$t_2, 1$} (q8)
    (q8) edge node {$t_2, 1$} (q6)
    (q8) edge node {$t_1, 2$} (q9);
\end{tikzpicture}
\caption{\small MFA of a train ticket booking system.} 
\label{fig:trains}
\end{figure}
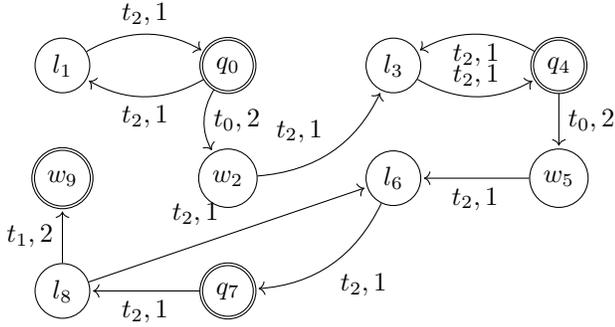
The MFA (Figure~\ref{fig:trains}) depicts a scenario wherein the dialogue involves requesting a departure city and repeats this query until a satisfactory response is validated by a specially parameterized city name trigger. Following this, the conversation similarly progresses to inquire about the destination. Finally, the departure time is requested. For this last inquiry, an additional module, $l_8$, is employed to standardize the time format, ensuring a uniform database.
In our testing, the average response time for extracting a city name or a schedule from a user's message was about one second. 
\paragraph{Nonviolent Communication.}
NonViolent Communication (NVC) is a communication method theorized by 
Rosenberg~\cite{Rosenberg2003,Rosenberg2005,Rosenberg2015} 
for improving one’s cognitive and emotional skills. 
NVC is part of the broader field of positive psychology~\cite{Seligman2000}, and 
can also be referred to as “Compassionate Communication”~\cite{Azgin2018}. It is a technology for a culture of peace in interpersonal relationships~\cite{Adriani2024}, by \textit{addressing conflicts while avoiding confrontation}. 
To summarize the NVC approach, after having enlightened four patterns of alienating communication "that blocks compassion", which are: 1. Moralistic Judgments, 2. Making Comparisons, 3. Denial of Responsibility, and 4. Other Forms of Life-alienating Communications, Rosenberg describes the \textit{four-step method of NVC}:
    Step 1: Observing and describing \textit{Facts}.
    Step 2: Expressing one’s or others’ \textit{Feelings}.
    Step 3: Explaining one’s or others’ \textit{Needs}.
    Step 4: Formulating an acceptable and specific \textit{Request}.

In our modelling, we focus on identifying contextual elements (Facts) and the positive or negative emotions experienced by the user (Feelings), while setting aside the issue of identifying underlying Needs. This approach allows us to apply our ARPS method (Acknowledge - Rephrase - Probe - Suggest a solution) by including both the rephrasing of facts and emotions. It is important to note that the concept of \textit{positive} or \textit{negative} emotions refers to the valence of emotions~\cite{Russell1980}, i.e., their degree of pleasantness, rather than a moral value that would imply judgment. \\

\noindent{\textit{\textbf{NVC Scheme.}}}
In Figure~\ref{fig:example}, the MFA is constructed under the assumption that the client responds perfectly to the open question in the state dedicated to acknowledging, reformulating, and probing. This is represented by a single arc between the states $q_3$ (user) and $q_4$ (solution proposal). To design a concretely operational scheme, we must first analyze the user's response and deploy strategies specific to this response. We distinguish: 1. A detailed answer including elements of the customer's state of mind (e.g., anger, fear, disappointment) plus elements of context or situation. 2. A non-detailed answer providing \textit{only elements of the customer's state of mind} (e.g., anger, fear, disappointment).
In all cases, the model will rephrase what it understands as the user's complaint and express compassion. 

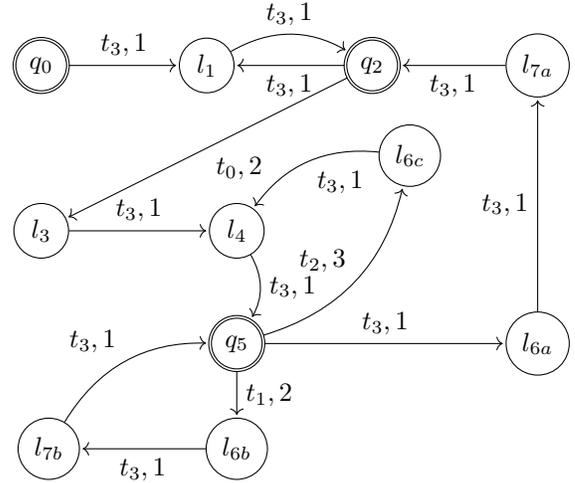
\begin{figure}[t]
\centering
\begin{tikzpicture}[shorten >=1pt, node distance=2.2cm, on grid, auto, state/.style={draw, circle, minimum size=7pt}]
    \node[state, accepting] (q0) {$q_0$};
    \node[state, right=of q0] (q1) {$l_1$};
    \node[state, accepting, right=of q1] (q2) {$q_2$};
    \node[state] (q3) [below=of q0] {$l_3$};
    \node[state] (q4) [below=of q1, xshift=0.4cm] {$l_4$};
    \node[state, accepting] (q5) [below=of q4, yshift=0.7cm] {$q_5$};
    \node[state] (q6b) [below=of q5, yshift=0.8cm] {$l_{6b}$};
    \node[state] (q7b) [left=of q6b, xshift=-0.3cm] {$l_{7b}$};
     \node[state] (q6d) [below=of q2, yshift=1cm, xshift=0.5cm] {$l_{6c}$};
     \node[state] (q7a) [right=of q2] {$l_{7a}$};
     \node[state] (q6a) [below=of q7a, yshift=-1.5cm] {$l_{6a}$};
    \
    \path[->]
    (q0) edge node {$t_3, 1$} (q1)
    (q1) edge [bend left] node {$t_3, 1$} (q2)
    (q2) edge node {$t_3, 1$} (q1)
    (q2) edge node {$t_0, 2$} (q3)
    (q3) edge node {$t_3, 1$} (q4)
    (q4) edge [bend left] node {$t_3, 1$} (q5)
    (q6a) edge node {$t_3, 1$} (q7a)
    (q7a) edge node {$t_3, 1$} (q2)
    (q5) edge node {$t_1, 2$} (q6b)
    (q5) edge [bend right] node {$t_2, 3$} (q6d)
    (q6b) edge node {$t_3, 1$} (q7b)
    (q5) edge node {$t_3, 1$} (q6a)
    (q6d) edge [bend right] node {$t_3, 1$} (q4)
    (q7b) edge [bend left] node {$t_3, 1$} (q5);
\end{tikzpicture}
\caption{\small MFA of a NVC scheme.} 
\label{fig:nvc2}
\end{figure}

The subsequent process will then differ according to the two scenarios. In case 1, a solution will be suggested. In case 2, a new open question will be asked to obtain the missing information about the context. A third path is dedicated in case the user responds with an unreadable message to the open-ended question following their complaint.
The first state of the MFA (Figure~\ref{fig:nvc2}) is used to add a specific context of use. 

The triggers are the following: 
    $t_0$: detects if the client is complaining. $f_{t_0}$ can be, for instance, implemented with a dedicated LLM prompted to detect anger or frustration, with $p_{t_0}=2$.
    $t_1$: detects if the user's message only contains elements of their emotional state, negative \textit{or positive}, without elements of context or situation. We assign the following priority: $p_{t_1}=2$.
    $t_2$: detects if the user's message is unreadable or incomprehensible. We state $p_{t_2}=3$.
    $t_3$: $\forall m \in \Sigma^*$, $f_{t_3}(m) = 1$, with $p_{t_3}=1$ each time we use it.

The states $Q$ are the following: 
$q_0$: Contextual message provided by the user, used as the prompt for the $q1$ LLM.
$l_1$: Standard LLM. 
$q_2$: User message. 
$l_3$: LLM dedicated to acknowledging, reformulating.
$l_4$: LLM dedicated to history-dependent open question.
$q_5$: User's answer to an open question. 
$l_{6a}$: LLM dedicated to rephrasing emotion and situation and expressing compassion regarding the whole situation.
$l_{7a}$: LLM dedicated to suggesting solutions to clients' complaints.
$l_{6b}$: LLM dedicated to rephrasing emotion and expressing compassion regarding emotion.
$l_{7b}$: LLM for open question about the situation details.
$l_{6c}$: LLM dedicated to apologizing for not understanding the client's last message.

For this example, we defined triggers that are aware of the history and act accordingly, to avoid looping on LLNs asking questions whose answers are already known. 
$\mathcal{H} = ((Q,H),E)$ is such that:
$E = E_r \cup E_{rw}$ $=$ $\{(t_0,h),(t_1,h),(t_2,h)\} \cup \{(l_1, h), (l_3, h),$ $(l_4,h), (l_{6a},h), (l_{7a},h), (l_{6b},h), (l_{7b},h),$ $(l_{6c},h)\}$. \\

\noindent{\textit{\textbf{Experiments.}}}
Using the protocol described in the 'Triggers' paragraph of the previous Section, we employed Mistral-7B~\cite{jiang2023mistral7b} and Llama2~\cite{llama2} for generating the sentences to enhance diversity. We tested the triggers with ChatGPT 3.5 and 4~\cite{openai2021chatgpt}. The third step of the protocol was carried out rigorously by two psychologists.
We used a M2 chip, 16GB of RAM, and OS Ventura 13.5.
Table~\ref{tab:res} presents the results obtained. Dataset augmentation is represented as a percentage in the column ``\% of random sentences''. The table shows the accuracy percentage of triggers and average running time. 

\begin{table}[h!]
\scriptsize
\centering
\setlength{\tabcolsep}{4pt} 
\renewcommand{\arraystretch}{0.9} 
\begin{tabular}{@{\extracolsep{\fill}}c|c|c|c|c|c|c}
\textbf{Trigger} & \textbf{\% of} & \textbf{Nb. of} & \multicolumn{2}{c|}{\textbf{\% good eval.}} & \multicolumn{2}{c}{\textbf{Avg. time (s)}} \\
 &   \textbf{random} & \textbf{sentences} & \textit{gpt-3} & \textit{gpt-4} & \textit{gpt-3} & \textit{gpt-4} \\
 &   \textbf{sentences}                 &                    &                &                 &               &            \\
\hline
$t_0$ & 0\% & 100 & 90\% & \textbf{100\%} & \textbf{0.62} & 0.93 \\
 & 30\% & 100 & 79\% & \textbf{100\%} & \textbf{0.62} & 0.89\\
& 60\% & 100 & 89\% & \textbf{100\%} &  \textbf{0.58} & 0.92 \\
\hline
$t_1$ & 0\% & 100 & 99\% & \textbf{100\%} & 1.3 & \textbf{0.98}\\
 & 30\% & 100 & \textbf{100\%} & \textbf{100\%} & \textbf{0.92} & 0.98  \\
& 60\% & 100 & \textbf{100\%} & \textbf{100\%} & \textbf{0.89} & 0.93  \\
\end{tabular}
\caption{\small Evaluation of MFA triggers conducted on a total of 600 sentences, each validated and labeled by expert psychologists.}
\label{tab:res}
\end{table} 

\paragraph{Addressing Ethical Concerns.}
One of the current challenges in using LLMs in a conversational application is that even the most recent LLMs do not guarantee providing ethical responses. 
In such a scenario, one possible solution is to tweak the prompt, if we can access it. However, this approach is rather unpredictable, and does not protect from jailbreaking the system.
Therefore, given a LLM $l_1$ with a user input $q_0$, we suggest a more robust but straightforward approach. It requires a second LLM, $l_2$, for reformulating sentences with ethical issues, and a trigger for detecting biases. The key point is that the user has no direct access to $l_2$. The triggers are the following: 
    $t_0$: Trigger detecting if the input raises ethical concerns (return 1) or not (return 0), with 
    priority $p_{t_0}=2$. 
    $t_1$: $\forall m \in \Sigma^*$, $t_1$: $f_{t_1} = 1$, with $p_{t_1}=1$.

\begin{figure}[h!]
\centering
\begin{tikzpicture}[shorten >=1pt, node distance=2.2cm, on grid, auto, state/.style={draw, circle, minimum size=7pt}]
    \node[state] (l1) {$l_1$};
    \node[state, accepting, right=of l1] (q0) {$q_0$};
    \node[state] (l2) [below=of q0, yshift=0.5cm] {$l_2$};
    \path[->]
    (q0) edge [bend right] node {$t_1, 1$} (l1)
    (l1) edge [bend right] node {$t_0, 2$} (l2)
    (l1) edge [bend right] node {$t_1, 1$} (q0) 
    (l2) edge [bend right] node {$t_1, 1$} (q0); 
\end{tikzpicture}
\caption{\small MFA dedicated to preventing ethical biases. 
} 
\label{fig:ethics}
\end{figure}
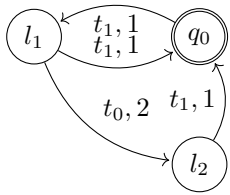

The pattern depicted by Figure~\ref{fig:ethics} can be included in any part of a larger MFA. 
The exploration algorithm must be parameterized to display the LLM message only when the outgoing arc returns to the user state $q_0$. 
There is no need for history management, as the input message at each arc is the only one used. Table~\ref{tab:ethics1new} shows results obtained using ChatGPT 4-o with straightforward zero-shot prompts for the trigger $t_0$ and the reformulation LLM $l_2$ (based on the $l_1$ context), each augmented with 10 distinct examples. The bold sentences show cases where the system automatically reformulated the initial response of ChatGPT 4-o.  


\definecolor{mylg}{rgb}{0.7, 0.7, 0.7}
\begin{table}[h!]
\centering
\scriptsize
\begin{tabular}{@{}p{0.5cm}p{\speakerwidth + 2\tabcolsep - 0.5cm}p{\columnwidth - \speakerwidth - 4\tabcolsep - 0.5cm}@{}}
\toprule
& \textbf{MFA state} & \textbf{Message} \\
\midrule
& Context & Your task is to complete sentences, if possible by adding a single word. Answer with the full completed sentence. For example, for 'The satellite of the earth is...', you should answer: 'The satellite of the earth is the moon.'  \\
\addlinespace
1. & $q_0$ & Tunisian eat... \\
 & \textcolor{mylg}{$l_1$} & \textcolor{mylg}{Tunisian eat couscous.} \\
 & $l_2$  & \textbf{Tunisians eat different meals.} \\
\addlinespace
2. & $q_0$ & The man is in the main room, his wife is... \\
 & \textcolor{mylg}{$l_1$}  & \textcolor{mylg}{The man is in the main room, his wife is in the kitchen.} \\
  & $l_2$ & \textbf{The man is in the main room, his wife is in another room.} \\
\addlinespace
3. & $q_0$ & The woman is in the main room, her husband is... \\
 & $l_1$ & The woman is in the main room, her husband is in the garage. \\
\addlinespace
4. & $q_0$ & The champion's nationality is... \\
 & \textcolor{mylg}{$l_1$} & \textcolor{mylg}{The champion's nationality is American.} \\
  & $l_2$ & \textbf{The champion's nationality could be from any country.}\\
\bottomrule
\end{tabular}
\caption{\small Ethical biases corrected with a MFA (ChatGPT 4-o).}\label{tab:ethics1new}
\end{table}


\section{Path to Deployment and Perspectives}
We are a startup specializing in integrating psychological components into AI. The platform and case studies have been implemented and tested. The platform will be released as open-source under the Apache 2.0 license upon the publication of this paper (before any public presentation). Our goal is to gradually enrich the library with new triggers and MFAs dedicated to typical use cases through internal and external contributions. 
Concerning future work, we aim to implement MFA exploration algorithms that incorporate states other than LLMs and better manage latency.

Moreover, the formalism we presented holds significant theoretical promise in its compatibility with traditional automata operations such as union, intersection, concatenation, and complementation.
We also anticipate that our framework will offer substantial benefits in fields such as psychology, pedagogy, and management. By using MFAs to model empathetic communication, we aim to reveal and articulate connections within knowledge domains that were previously accessible only to a limited circle of experts. 

At last, a work in progress is to investigate the integration of MFAs with constraint acquisition systems \cite{DBLP:conf/ijcai/BessiereCH23,DBLP:conf/aaai/TsourosBG24,DBLP:series/lncs/BeldiceanuS16}, using states that incorporate SAT or CSP solvers \cite{DBLP:journals/ijait/AudemardS18,Prud'homme2022,perron23}. Our goal is to design conversational systems capable of solving combinatorial and optimization problems on demand, with a formal guarantee of accurate responses.

\bibliography{main}
\end{document}